\documentclass[aip,jcp,reprint,graphicx]{revtex4-1}
\usepackage{graphicx}
\usepackage{amsmath}
\usepackage{color}
\usepackage{times} 
\usepackage{dcolumn}
\usepackage{bm}

\newcommand{\dr}{^{\circ}}

\newcommand{\SSE}{\langle\hat{S}^{\,2}\rangle}
\newcommand{\tr}{\mbox{Tr}}

\newcommand{\be}{\begin{equation}}
\newcommand{\ee}{\end{equation}}
\newcommand{\bea}{\begin{eqnarray}}
\newcommand{\eea}{\end{eqnarray}}

\newcommand{\ie}{{\it i.e.,\ }}

\newcommand{\hh}{H$_4$ }

\draft

\begin{document}

\title{H$_4$: A Challenging System For Natural Orbital Functional Approximations}

\author{Eloy Ramos-Cordoba}
\affiliation{Faculty of Chemistry, University of the Basque Country UPV/EHU,
and Donostia International Physics Center (DIPC). 
P.K. 1072, 20080 Donostia, Euskadi, Spain}
\email{eloy.raco@gmail.com, ematito@gmail.com}
\author{Xabier Lopez}%

\affiliation{Faculty of Chemistry, University of the Basque Country UPV/EHU,
and Donostia International Physics Center (DIPC). 
P.K. 1072, 20080 Donostia, Euskadi, Spain}
\author{Mario Piris}%

\affiliation{Faculty of Chemistry, University of the Basque Country UPV/EHU,
and Donostia International Physics Center (DIPC). 
P.K. 1072, 20080 Donostia, Euskadi, Spain}
\affiliation{IKERBASQUE, Basque Foundation for Science, 48011 Bilbao, Spain}
\author{Eduard Matito}%

\affiliation{Faculty of Chemistry, University of the Basque Country UPV/EHU,
and Donostia International Physics Center (DIPC). 
P.K. 1072, 20080 Donostia, Euskadi, Spain}
\affiliation{IKERBASQUE, Basque Foundation for Science, 48011 Bilbao, Spain}

\date{\today}

\begin{abstract}
 The correct description of nondynamic correlation by electronic structure methods 
not belonging to the multireference family is a challenging issue. The transition
of $D_{2h}$ to $D_{4h}$ symmetry in \hh molecule is among the most simple archetypal 
examples to illustrate the consequences of missing nondynamic correlation effects. 
The resurge of interest in density matrix functional methods has brought several
new methods including the family of Piris Natural Orbital Functionals (PNOF).
In this work we compare PNOF5 and PNOF6, which include nondynamic electron correlation
effects to some extent, with other standard ab initio methods in the \hh $D_{4h}/D_{2h}$ 
potential energy surface. 
Thus far, the wrongful behavior of single-reference methods at the $D_{2h}-D_{4h}$ transition of \hh
has been attributed to wrong account of nondynamic correlation effects, whereas
in geminal-based approaches it has been assigned to a wrong coupling of spins 
and the localized nature of the orbitals.
We will show that actually \textit{interpair} nondynamic correlation is the key to a 
cusp-free qualitatively correct description of \hh PES. By introducing \textit{interpair} nondynamic correlation,
PNOF6 is shown to avoid cusps and provide the correct smooth PES features at distances close to the equilibrium, total and local spin 
properties along with the correct electron delocalization,  as reflected by natural orbitals
and multicenter delocalization indices. 
\end{abstract}


\maketitle

\section{Introduction}

The correct description of nondynamic correlation effects is a challenging 
task for electronic structure methods. In wave function approaches, a multireference ansatz 
is needed to properly account for these effects. The computational scaling cost of 
such methods limits their use to systems of moderate size. Within density functional theory (DFT) the 
proper inclusion of nondynamic correlation effects is an open problem. \cite{becke_perspective:_2014}
In practice, a broken-symmetry calculation is usually performed producing 
wrong spin densities.~\cite{bulik_can_2015} \\

 An alternative to both wave function and DFT methods is natural orbital functional theory (NOFT).\cite{Gilbert1975,Piris2007,pernal:15tcc}
 In recent years, 
several functionals have been proposed by reconstruction of the two-particle reduced density matrix (2-RDM) in terms 
of the one-particle reduced density matrix (1-RDM).\citep{Ludena2013} In particular, within the family of Piris Natural Orbital 
Functionals (PNOF),~\cite{piris:13ijqc,piris:14ijqc} PNOF5~\cite{piris:11jcp} and PNOF6~\cite{piris:14jcp} are among the best candidates to treat nondynamic 
correlated systems. They describe properly the dissociation limit of several molecules, recovering
the correct integer number of electrons on each fragment upon dissociation.\cite{matxain:11pccp,piris:14jcp}
Both PNOF5 and PNOF6 belong to the family of orbital-pairing approaches, but the former only 
includes intrapair electron correlation while in the latter electrons on different pairs are also 
correlated. 
The inclusion of interpair electron correlation in PNOF6 allows a better description of 
correlation effects and it also removes the symmetry-breaking artifacts that are present in independent-pairs
approaches such as PNOF5 when treating delocalized systems.~\cite{piris:14jcp}\newline

The purpose of this manuscript is to analyze the effect of interpair electron 
correlation on the treatment 
of nondynamic correlation by investigating the performance of PNOF5 and PNOF6
and several standard \textit{ab initio} computational methods. 
To this end we will examine the $D_{4h}/D_{2h}$
potential energy surface of the planar \hh model (hereafter, simply PES).  \newline

\hh has been extensively used to test single-reference post-Hartree-Fock methods~\cite{robinson_application_2012,bulik_can_2015,
voorhis_benchmark_2000,jankowski_physical_1999-1,jankowski_physical_1999,kowalski_towards_1998,kowalski_full_1998,
paldus_application_1993,robinson_quasi-variational_2012,robinson_benchmark_2012,sand_parametric_2013}
and geminal-based theories.~\cite{jeszenszki_local_2015,Rassolov2007a}
Hartree-Fock, MP2 and MP3 show a spurious cusp on the PES of \hh as the system evolves from
$D_{2h}$ to $D_{4h}$ symmetry. The cusp is the maximum energy value along the symmetry transition. Conversely, traditional coupled cluster (CC) methods predict a cusp but this cusp is a
local miminum in the $D_{2h}-D_{4h}$ transition.
Recently, Bulik \textit{et al.} have shown that an improvement of the 
description of correlated systems can be also achieved by removing terms in traditional CC theory.~\cite{bulik_can_2015}
Variational CC approaches also improve this wrong behavior of the traditional CC implementations,\cite{voorhis_benchmark_2000,robinson_quasi-variational_2012,robinson_application_2012}
however, most of these approaches
revert the local minimum to a local maximum but most of them do not avoid the
presence of a spurious cusp.
Geminal-based theories predict a (maximum) cusp at the square geometry. Jeszenszki \textit{et al.}\cite{jeszenszki_local_2015} 
have attributed this failure to an insufficient account of spin couplings and the localized character of the orbitals. By including triplet components in the geminals, the orbitals become delocalized and the characteristic cusp vanishes, but the resultant PES is not completely smooth and wave function becomes spin contaminated. 
The authors also examined the local 
spin\cite{ramos-cordoba_toward_2012,ramos-cordoba_local_2012,ramos-cordoba_two_2014} of the system using different geminal-based approaches. 
Jeszenszki \textit{et al.}\cite{jeszenszki_local_2015} have found that singlet-coupled geminals fail to describe correctly local spins at the D$_{4h}$ geometry. The inclusion of triplet 
components improve the results but the local spin values are not smooth along the PES.\\

Thus far, the wrongful behavior of single-reference methods at the $D_{2h}-D_{4h}$ transition of \hh
has been ascribed to a wrong account of nondynamic correlation effects,\cite{robinson_application_2012,bulik_can_2015,
voorhis_benchmark_2000,robinson_quasi-variational_2012} whereas
in geminal-based approaches, the spurious (maximum) cusp has been attributed to a wrong coupling of spins 
and the localized nature of the orbitals.\cite{jeszenszki_local_2015}
We will show that actually \textit{interpair} nondynamic correlation is the key to qualitative
cusp-free correct description of \hh PES. By introducing \textit{interpair} nondynamic correlation,
PNOF6 is shown to avoid cusps and provide the correct smooth PES features, total and local spin 
properties along with the correct electron delocalization, as reflected by natural orbitals
and multicenter delocalization indices. 

\section{Theory}

\subsection{PNOF5/PNOF6}

 In this section we will briefly review the formulation of PNOF5~\cite{piris:11jcp} and PNOF6.~\cite{piris:14jcp} 
 Both PNOF5 and PNOF6 
belong to the family of orbital-pairing methods, which divide the spatial orbital space into
subspaces (a set of orbitals) that contain two electron each.
These methods couple each orbital $g$
below the Fermi level ($F=N/2$, where $N$ is the number of electrons of the system) 
with $N_c$ orbitals above it, being 
$\Omega_{g}$ the 
subspace containing orbital $g$ and its coupled counterparts. 
The original formulations of both functionals were introduced for $N_c=1$ but 
subsequently extended versions
($N_c>1$) were reported.~\cite{Piris2013e,lopez:15jpca}
The sum rule for the occupation numbers ($n$) is fulfilled for each of the $N/2$ subspaces $\Omega_{g}$, 

\begin{equation}\label{eq:pairing}
\sum_{p\in\Omega_{g}}n_{p}=1
\end{equation}
where $p$ denotes a spatial natural orbital (NO) and $n_p$ its occupation number. \\

The PNOF5 and PNOF6 energy expressions for a singlet state system can be written as
\begin{equation}
E=\sum\limits _{g=1}^{F}E_{g}+\sum\limits _{f\neq g}^{F}\sum\limits _{p\in\Omega_{f}}\sum\limits _{q\in\Omega_{g}}E_{pq}^{int}.\label{PNOF56}
\end{equation}
The first term of Eq. (\ref{PNOF56}) corresponds to the sum of energies of 
$F$ independent pairs with energy $E_{g}$, namely,

\begin{equation}
E_{g}=\sum\limits _{p\in\Omega_{g}}n_{p}\left(2\mathcal{H}_{pp}+\mathcal{J}_{pp}\right)+\sum\limits _{p,q\in\Omega_{g},p\neq q}E_{pq}^{int},\label{Eg}
\end{equation}
where $\mathcal{H}_{pp}$ is the matrix element of the kinetic energy
plus nuclear-electron attraction terms and $\mathcal{J}_{pp} = \left<pp|pp\right>$
is the Coulomb interaction between two electrons with opposite spins
at the spatial orbital $p$.
The term $E_{pq}^{int}$ contains the interaction energy between electrons 
in different spatial orbitals $p$, and $q$,
\begin{equation}
E_{pq}^{int}=\left(n_{q}n_{p}-\Delta_{qp}\right)\left(2\mathcal{J}_{pq}-\mathcal{K}_{pq}\right)+\Pi_{qp}\mathcal{L}_{pq}\label{Eint}
\end{equation}
where $\mathcal{J}_{pq}=\left\langle pq|pq\right\rangle $ and $\mathcal{K}_{pq}=\left\langle pq|qp\right\rangle $
are the direct and exchange integrals, respectively and $\mathcal{L}_{pq}=\left\langle pp|qq\right\rangle $
is the exchange and time-inversion integral.\cite{piris:99jmc} Matrices $\Delta$ and $\Pi$ are auxiliary matrices 
proposed~\cite{piris:06ijqc} to reconstruct the 2-RDM in terms of the NO occupancies. The diagonal elements of these matrices are $\Delta_{pp}=n_p^2$ and $\Pi_{pp}=n_p$. The off-diagonal 
elements of $\Delta$ and $\Pi$ determine the different implementation of the PNOF$i$ $(i=1-6)$ series.
In particular, PNOF5 and PNOF6 differ on the treatment of the interaction
between electrons on different pairs.\newline

In PNOF5, when orbitals $p$ and $q$ belong to the same subspace $\Omega_g$, the off-diagonal elements of $\Delta$ and $\Pi$ are $\Delta_{pq}=n_qn_p$ and 
\begin{equation}
\begin{array}{c}
 \Pi_{pq}=\begin{cases}
-\sqrt{n_qn_p}\,, & p = g \quad or \quad q = g\\
\sqrt{n_qn_p}\,, & p,q>F,
\end{cases}
\end{array}
\end{equation}\newline
respectively, and they vanish when $p$ and $q$ belong to different subspaces.
Consequently, the second term of Eq. \ref{PNOF56} becomes 
\begin{equation}
\sum\limits _{f\neq g}^{F}\sum\limits _{p\in\Omega_{f}}\sum\limits _{q\in\Omega_{g}}E_{pq}^{int}(\text{PNOF5})=n_{q}n_{p}\left(2\mathcal{J}_{pq}-\mathcal{K}_{pq}\right).\label{Eintpnof5}
\end{equation}
The expression above indicates that the interaction between electrons in different pairs is 
treated at the mean-field level. 
Therefore, PNOF5 lacks correlation between electrons in different pairs. 
In contrast, the PNOF6 $\Delta_{pq}$ and $\Pi_{pq}$ matrices
(when $p$ and $q$ belong to different subspaces these matrices do not vanish) 
include terms that account for interpair electron correlation.  The off-diagonal
elements $\Delta_{pq}$ and $\Pi_{pq}$ in PNOF6 read as 
\begin{equation}
\begin{array}{cc|cc|cc}
\Delta_{qp} &  & \Pi_{qp} &  &  & Orbitals\\
\hline e^{-2S}h_{q}h_{p} &  & -e^{-S}\left(h_{q}h_{p}\right)^{\frac{1}{2}} &  &  & q\leq F,p\leq F\\
\frac{\gamma_{q}\gamma_{p}}{S_{\gamma}} &  & -\Pi_{qp}^{\gamma} &  &  & \begin{array}{c}
q\leq F,p>F\\
q>F,p\leq F
\end{array}\\
e^{-2S}n_{q}n_{p} &  & e^{-S}\left(n_{q}n_{p}\right)^{\frac{1}{2}} &  &  & q>F,p>F
\end{array}\label{DPi}
\end{equation}
where $h_{p}$ is the hole $(1-n_{p})$ in the spatial orbital
$p$ and $S$, $\gamma_p$, $S_{\gamma}$, and $\Pi^{\gamma}$ are 
defined as
\begin{equation}
\begin{array}{c}
S={\displaystyle \sum_{q=1}^{F}}h_{q},\quad \alpha_{p}=\begin{cases}
e^{-S}h_{p}\,, & p\leq F\\
e^{-S}n_{p}\,, & p>F
\end{cases}\\
S_{\alpha}={\displaystyle \sum_{q=1}^{F}}\alpha_{q}, \quad \gamma_{p}=n_{p}h_{p}+\alpha_{p}^{2}-\alpha_{p}S_{\alpha}\\
\\
S_{\gamma}={\displaystyle \sum_{q=1}^{F}}\gamma_{q}, \quad
\Pi_{qp}^{\gamma}=\left(n_{q}h_{p}+\frac{\gamma_{q}\gamma_{p}}{S_{\gamma}}\right)^{\frac{1}{2}}\left(h_{q}n_{p}+\frac{\gamma_{q}\gamma_{p}}{S_{\gamma}}\right)^{\frac{1}{2}}
\end{array}
\end{equation}\newline

Recently, PNOF5 has been proved equivalent to an antisymmetrized product of 
strongly orthogonal geminals (APSG).\cite{pernal:13ctc,Piris2013c} Conversely,
PNOF6 is not related to geminal theories but it keeps the orbital-pairing
scheme, Eq.~\ref{eq:pairing}. In this work we have used
 the $N_c=1$ version of the functionals. That is, each orbital subspace contains two spatial 
 orbitals and then only $N$ spatial orbitals are correlated. 
In this sense, both functionals take into account most of the nondynamic correlation effects, but 
while PNOF5 includes only intrapair correlation, PNOF6 incorporates also the interpair correlation, through $\Delta$ and $\Pi$ matrices defined in Eq. \ref{DPi} (see Eq.\ref{Eint})

\subsection{Local Spin And Electron Delocalization}

Local spins can be obtained by decomposing the expectation value of the total spin square operator $\SSE$
into atomic  or fragment contributions as 
\be
		\SSE = \sum_A \SSE_A + 
		\sum_{A\neq B} \SSE_{AB},
\ee 
where $\SSE_A$ is the local spin on fragment $A$ and $\SSE_{AB}$ accounts for the coupling between spins on fragments $A$ and $B$. Recently some of us have presented a general formulation of the local spin that fulfills a set of physical constrains.\cite{ramos-cordoba_toward_2012,ramos-cordoba_local_2012} For singlet systems, the formulation reads as

\begin{eqnarray}
\label{1ccomp}
\SSE_{A} &&= \frac{3}{4}\left( 2\, \tr({^1\bf DS}^{A}) - \tr({^1\bf DS}^{A}{^1\bf D}) \right)\\ \nonumber
 &&+ \frac{1}{2}\sum_{ijkl}\Gamma_{ij;kl}S^A_{ki}S^A_{lj}-\frac{1}{2}\sum_{ijkl}\Gamma_{ij;kl}S^A_{li}S^A_{kj} 
\end{eqnarray}
and 
\begin{eqnarray}
\label{2ccomp}
\SSE_{AB} =
 \frac{1}{2}\sum_{ijkl}\Gamma_{ij;kl}S^A_{ki}S^B_{lj}-\frac{1}{2}\sum_{ijkl}\Gamma_{ij;kl}S^A_{li}S^B_{kj}
\end{eqnarray}

where ${^1\bf D}$, $\Gamma$, and $S^A$ are the spinless 1-RDM, the spinless cumulant of the 2-RDM, and the fragment 
orbital overlap matrix.\cite{ramos-cordoba_toward_2012}
The correct description of local spins has been recently put forward as a stringent condition to test natural-orbital based cumulant matrix (or 2-RDM) approximations,\cite{ramos-cordoba_two_2014} and has been used to characterize and quantify the diradical and triradical character of molecules.\cite{ramos2014characterization,ramos2014diradical} In this work, we will use the local spin analysis to study the effect of the interpair electron correlation in PNOF5 and PNOF6 on the spin coupling of electrons located at different atoms.\newline

The calculation of electron delocalization among different fragments can be performed
through the NO-weighted overlap multiplications involving the different fragments.
This is commonly known as Giambiagi's multicenter index~\cite{giambiagi:90sc}
and its expression reads~\cite{cioslowski:07jpca}
\begin{equation}\label{eq:multi}
I_{\textit{ABCD}}=\sum_{ijkl}n_in_jn_kn_lS_{ij}^AS_{jk}^BS_{kl}^CS_{li}^D
\end{equation}
The quantity has been successfully used to account for several multicenter delocalization
phenomena including multicenter bonding,~\cite{feixas:14jctc} conjugation effects~\cite{feixas:11jpca} 
and aromaticity.~\cite{giambiagi:00pccp,feixas:15csr}

\section{Computational Details}
In this work we have computed the D$_{4h}$/D$_{2h}$ PES of \hh employing the following
methods:
Hartree-Fock (HF), CC singles and doubles (CCSD), 
CCSD with perturbative estimation of triple excitations (CCSD(T)), complete active space self-consistent field CASSCF (with a 4 electrons in 4 orbitals active space), PNOF5, PNOF6 and full configuration interaction (FCI). 
This benchmark data set includes methods that mostly include dynamic correlation
effects (CCSD and CCSD(T)) or nondynamic correlation effects (CASSCF) and will be used
as benchmark references to measure the amount of dynamic and nondynamic correlation
effects included in PNOF5 and PNOF6.\newline

All calculations based on wave function methods have been performed with the Gaussian03 
\cite{g03c02_short} 
set of programs except those at the FCI level that were performed with a modified version 
of the program of Knowles and Handy.\cite{knowles_new_1984,knowles_determinant_1989}
NOF calculations have been carried out using DoNOF program. The matrix elements of the 
kinetic energy, the nuclear-electron attraction energies, and the one- and two-electron integrals needed to perform the PNOF calculations  
have been obtained from GAMESS.
\cite{schmidt_general_1993,gordon_chapter_2005}
The correlation-consistent aug-cc-pVDZ \citep{jr__1989} basis set has been employed for all the calculations. The local spin analysis has been performed using DMN\cite{dmn} to compute the 2-RDM and APOST-3D\cite{apost} to 
calculate the local spins using the topological fuzzy Voronoi cells to define the atomic regions.\cite{salvador_communication:_2013} \\

\section{Results}

\begin{figure}[h!]
 \includegraphics[scale=0.6]{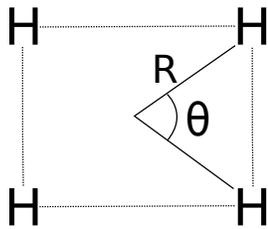}
 \caption{ D$_{4h}$/D$_{2h}$ \hh model.}
 \label{model}
\end{figure}

The PES 
of \hh is characterized using two parameters, R and $\theta$ (see Fig. \ref{model}). The former, controls 
the distance between each H atom and the center of mass while the latter measures the angle formed by two neighbor H 
atoms and the center of mass (see Fig. \ref{model}).
At $\theta=90\dr$, the system possesses D$_{4h}$ symmetry and two configurations 
with symmetries $a^2_gb^2_{2u}$ and $a^2_gb^2_{3u}$ become degenerate. By modifying $\theta$ one can control the degree of symmetry
distortion with respect to the D$_{4h}$ ($\theta=90\dr$) structure, thus modulating the multireference character (and hence the nondynamic correlation) of the system. 
In this sense, the \hh PES represents a challenging system for most electronic structure methods 
as it combines nondynamic correlation and dynamic correlation effects.\newline

\begin{figure*}
 \includegraphics[scale=0.90]{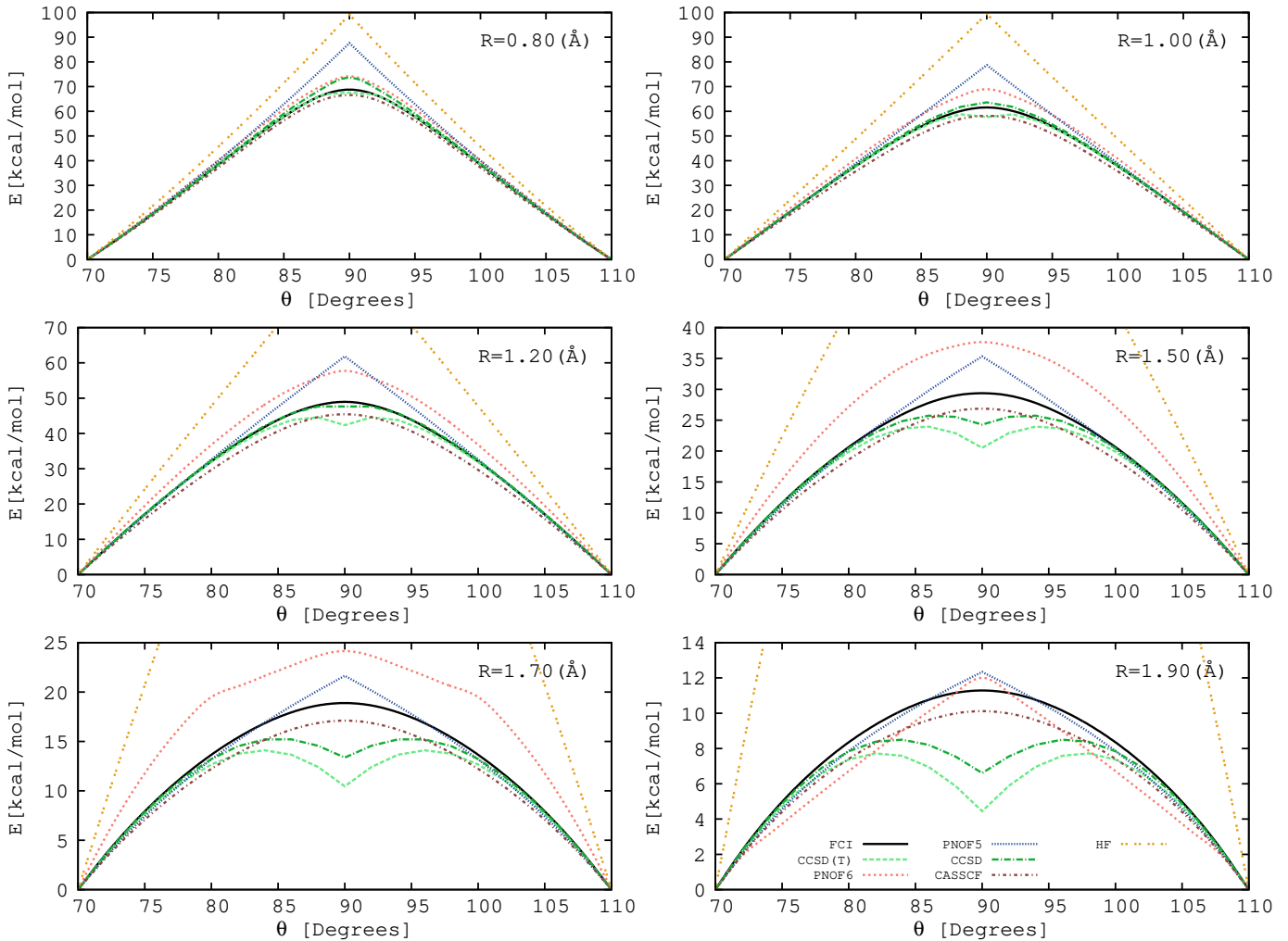}%

 \caption{\label{R2} Relative energies in kcal/mol with respect to the lowest energy found for each method $\theta=70\dr$, along the $D_{2h}/D_{4h}$ PES of \hh. }
\end{figure*}

The relative energies with respect to the minimum energy at $\theta=70\dr$ for each method of the \hh model keeping $R$ constant 
for different distances and modifying $\theta$ are shown in Fig.~\ref{R2}. The system is symmetric at $\theta=90\dr$
and it is described by two degenerate configurations, which correspond to the minimum HF solutions at 
$\theta<90\dr$ and $\theta>90\dr$, respectively.
The FCI curve has an energy maximum at $\theta=90\dr$ and the energy curve is smooth 
along the entire range of angles. The energy needed to change from $\theta=70\dr$ to the 
D$_{4h}$ geometry 
decreases gradually as the radius $R$ increases until 
the PES becomes considerably flat. 
The CASSCF curves show the right qualitatively features, \ie a maximum at $\theta=90\dr$ 
and a smooth transition from $\theta=70\dr$ to $\theta=110\dr$.
However, due to missing dynamic correlation energy that becomes important at the $\theta\gg90\dr$ and $\theta\ll90\dr$ regions, CASSCF relative values are downshifted 
to lower energies.\newline

At $\theta=90\dr$ two configurations become degenerate and the HF solution presents symmetry-breaking artifacts that result in a maximum cusp in the energy profile.~\cite{kowalski_full_1998}
Therefore, it is only natural that most post-HF single-reference methods based on the 
RHF reference also fail to qualitatively describe this PES. 
Although at small $R$ values CCSD and CCSD(T) 
mimic the FCI PES, as the radius $R$ increases 
first CCSD(T) (at $R=0.80 \AA$) and then CCSD (at $R=1.20 \AA$) break down and show
a cusp of the PES at $\theta=90\dr$, which ---unlike the HF cusp--- is a local minimum
with respect to $\theta$. Since CASSCF with a (4,4) active space shows a qualitative right result and dynamic-correlation-including
methods produce an artifact at $\theta=90\dr$, one attributes this feature to 
the lack of nondynamic correlation effects. 
Consequently, at short values of $R$ and for the $\theta$ values considered, the CC results 
are in perfect agreement with FCI. 
\newline

PNOF5 ---a nondynamic-correlation-including method--- shows a maximum cusp at $\theta=90\dr$,
like VCC,~\cite{voorhis_benchmark_2000} OQVCCD and OQVCCD(T),~\cite{robinson_benchmark_2012} 
and the lately introduced CCD0 and CCSD0, which are single-reference CC variants that exclude
certain excitations.~\cite{bulik_can_2015}
This result suggests that PNOF5 is missing some nondynamic correlation and it is only this
fraction of nondynamic correlation that is responsible for the spurious cusp. 

\begin{figure*}
 \includegraphics[scale=0.90]{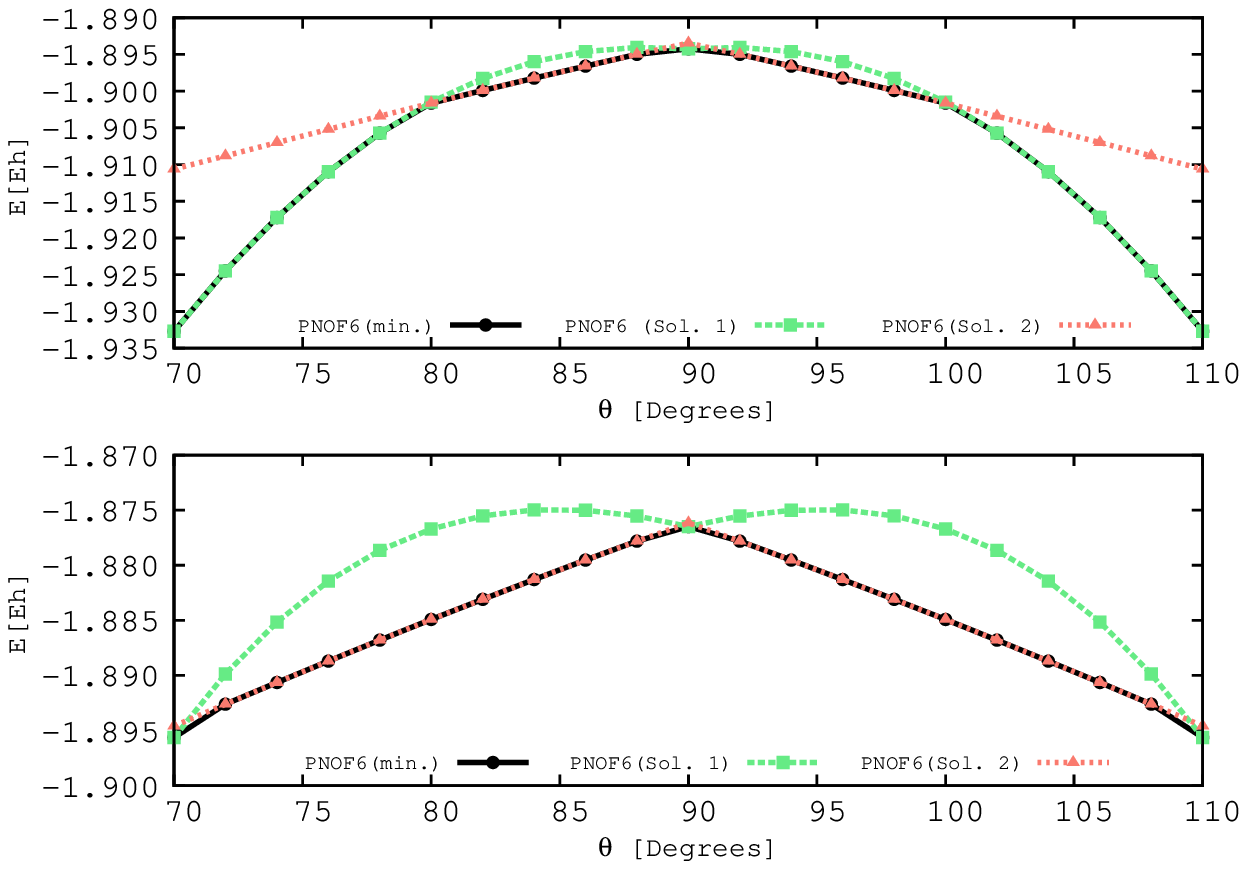}%

 \caption{ Absolute PNOF6 energies in a.u. for  $R=1.70 \AA$ (top) and  $R=1.90\AA$(bottom). PNOF6(Sol. 1 ) and PNOF6(Sol. 2 ) stand for the two solutions that show a crossing and PNOF6 (min.) stand for the minim energy solution of each value of $\theta$.}
 \label{abs-170-190}
\end{figure*}

On the other hand, PNOF6 which ---at variance with PNOF5--- includes 
interpair correlation, shows a smooth PES for $R\leq1.5\AA$, suggesting that only interpair
nondynamic correlation is actually needed to obtain a cusp-free, qualitatively correct description
of the \hh PES at values close to the minimum energy geometry. 
When $R=1.70\AA$ and $1.90\AA$, the PNOF6 solution is not perfectly smooth. This behavior is due to the crossing of two solutions of the PNOF6 equations as can be seen in Fig. \ref{abs-170-190}. In this graphic, the minimum PNOF6 solution is showed in solid lines. One can see the crossing of two solutions at $\theta\simeq80 \dr , 90\dr,$ and $100\dr$ for $R=1.70\AA$ and at $\theta\simeq70\dr, 90\dr$, and $110\dr$ for $R=1.90\AA$. 
At large $R$ only one solution (labeled Sol. 2 in Fig. \ref{abs-170-190}) of the PNOF6 equations is found, there is no longer a crossing and the PES smoothness is recovered, the shape of PNOF6 and FCI relative energies being almost indistinguishable.\newline

Table \ref{table-1} gathers the relative energies at
$\theta= 70 \dr$ with respect to the energy at $\theta= 90 \dr$.  
\begin{table}
\centering
\caption{Relative energies (kcal/mol) as the difference between the absolute energies at $\theta=90\dr$ and $\theta=70\dr$
for different values of $R (\AA)$.}
\label{table-1}
\begin{tabular}{ccccccc}\hline
Method & R=0.80 & R=1.00 & R=1.20 & R=1.40 & R=1.60 & R=1.80  \\ \hline
FCI    & 68.75  & 61.54  & 48.92  & 35.58  & 23.79  & 14.72 \\
HF    & 99.15  & 99.43  & 93.38  & 85.20  & 76.59  & 68.33\\
CASSCF & 66.61  & 58.19  & 45.44  & 32.68  & 21.66  & 13.27 \\
PNOF5  & 87.63  & 78.67  & 61.88  & 43.67  & 27.94  & 16.47 \\
PNOF6  & 74.19  & 68.98  & 57.74  & 44.48  & 30.68  & 18.17 \\\hline
\end{tabular}
\end{table}   
For $R=0.8\;\AA$, $R=1.00\;\AA$, and $R=1.20\;\AA$ PNOF6 improves PNOF5 (as to compared to FCI) by 
13.44, 9.69 and 4.14 kcal/mol, respectively. 
At larger values of $R$, PNOF5 improves over PNOF6 but the difference between them does 
not exceed 3 kcal/mol. CASSCF results
 are closer to FCI than PNOF6 for all the distances. The difference attains its maximum 
at $R=1.20\;\AA$, in which CASSCF is 12.29 kcal/mol closer
 to FCI than PNOF6. 
These deviations put forward the current limits of PNOF6 to fully account 
for correlation effects.\newline

In table \ref{abs-tab} we collect FCI, PNOF5, and PNOF6 absolute energies for $R=0.80\AA$, $1.20\AA$, and $1.70\AA$. PNOF5 energies are in all cases closer to FCI than PNOF6. This is
due to the repulsive electron-electron interpair correlation energy term  that is included in the PNOF6 functional. PNOF6 improves 
qualitatively the shape of the PES, provides good relative energies at the price of higher absolute energies.   \newline
\begin{table*}

\centering
\caption{FCI, PNOF5, and PNOF6 \hh absolute energies in a.u. for different values of $\theta$ and $R$ }
\label{abs-tab}
\begin{tabular}{c|ccc|ccc|ccc} \hline
   &    \multicolumn{3}{ c |}{$R=0.80\;\AA$ }      &     \multicolumn{3}{ c| }{$R=1.20\;\AA$ }         &    \multicolumn{3}{ c }{$R=1.70\;\AA$ } \\
$\theta\dr$  & FCI      & PNOF5    & PNOF6    & FCI      & PNOF5    & PNOF6    & FCI      & PNOF5    & PNOF6    \\ \hline
70     & -2.20639 & -2.16625 & -2.14177 & -2.14307 & -2.11972 & -2.06829 & -2.04310 & -2.03505 & -1.93271  \\
72     & -2.19498 & -2.15467 & -2.12979 & -2.13184 & -2.10876 & -2.05554 & -2.03759 & -2.02993 & -1.92448 \\
74     & -2.18305 & -2.14247 & -2.11728 & -2.12105 & -2.09810 & -2.04323 & -2.03271 & -2.02533 & -1.91722  \\
76     & -2.17071 & -2.12970 & -2.10437 & -2.11077 & -2.08772 & -2.03143 & -2.02840 & -2.02118 & -1.91097 \\
78     & -2.15805 & -2.11639 & -2.09113 & -2.10104 & -2.07760 & -2.02022 & -2.02462 & -2.01744 & -1.90572 \\
80     & -2.14523 & -2.10259 & -2.07771 & -2.09195 & -2.06773 & -2.00970 & -2.02135 & -2.01404 & -1.90165  \\
82     & -2.13244 & -2.08832 & -2.06428 & -2.08364 & -2.05809 & -2.00005 & -2.01855 & -2.01094 &  -1.89992 \\
84     & -2.12005 & -2.07358 & -2.05117 & -2.07635 & -2.04864 & -1.99151 & -2.01626 & -2.00809 &  -1.89823 \\
86     & -2.10881 & -2.05839 & -2.03902 & -2.07046 & -2.03936 & -1.98445 & -2.01451 & -2.00543 &  -1.89659 \\
88     & -2.10023 & -2.04273 & -2.02916 & -2.06652 & -2.03019 & -1.97926 & -2.01339 & -2.00294 &  -1.89499 \\
90     & -2.09683 & -2.02660 & -2.02354 & -2.06512 & -2.02111 & -1.97628 & -2.01300 & -2.00055 & -1.89423 \\\hline
\end{tabular}

\end{table*}

\begin{figure}
 \includegraphics[scale=0.30]{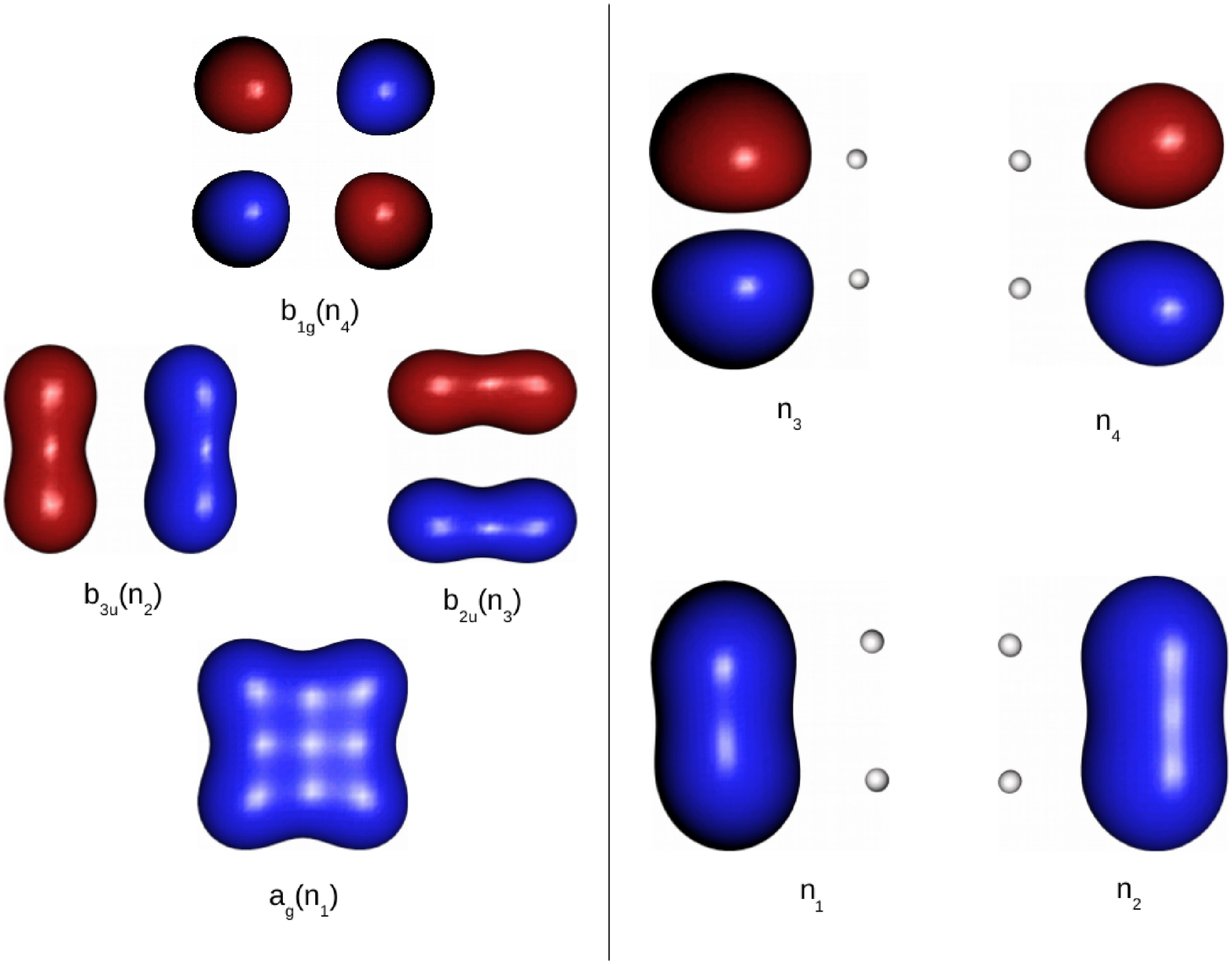}%

 \caption{\label{R1} PNOF6 (left), and PNOF5 (right) natural orbitals of \hh for $R=1.0\;\AA$ and $\theta=90\dr$}
\end{figure}

APSG, which is the antisymmetric
wavefunction behind PNOF5,~\cite{pernal:13ctc} has been shown to also exhibit this spurious
maximum cusp at $\theta=90\dr$.~\cite{Rassolov2007a}
The failure of APSG has been attributed to the localized nature of its orbitals and the wrong 
account of spin coupling.
Szabados and coworkers~\cite{jeszenszki_local_2015} have demonstrated that APSG using delocalized
orbitals, which correspond to a solution of the ASPG equations, eliminates the cusp. 
In Fig. \ref{R1} we plot the orbitals that arise from PNOF6 and PNOF5 at $R=1.0\;\AA$ and $\theta=90\dr$.
PNOF5 NO are localized on $H-H$ bonds and each bonding orbital is coupled with 
its antibonding counterpart. At this value of $\theta$, the same picture with the 
orbitals \textit{horizontally} localized is equivalent. On the other hand, the PNOF6 NO present 
the expected delocalized character and mimic the canonic orbitals obtained in a HF calculation. Importantly, both solutions showed in Fig. \ref{abs-170-190} for $R=1.70\AA$ and $1.90\AA$ present delocalized orbitals. Unlike PNOF5, PNOF6 equations do not lead to a stationary solution that corresponds to a set of localized orbitals.

\begin{table*}
\centering

\centering
\caption{CASSCF(4,4), PNOF5, and PNOF6 NO occupation numbers at $\theta=90\dr$ for different values of $R$.}
\label{occ}
\begin{tabular}{ccccc} \hline
$R (\AA)$    & $n_1$     & $n_2$    & $n_3$    & $n_4$    \\ \hline
     & \multicolumn{4}{ c }{CASSCF} \\
0.80 & 1.939  & 1.000 & 1.000 & 0.061 \\
1.00 & 1.882  & 1.000 & 1.000 & 0.118 \\
1.20 & 1.795  & 1.000 & 1.000 & 0.205 \\
1.50 & 1.604  & 1.000 & 1.000 & 0.396 \\
1.70 & 1.458  & 1.000 & 1.000 & 0.542 \\
1.90 & 1.327  & 1.000 & 1.000 & 0.673 \\
20.00 & 1.000  & 1.000 & 1.000 & 1.000 \\
     & \multicolumn{4}{ c }{PNOF5}      \\
0.80 & 1.923  & 1.921 & 0.079 & 0.077 \\
1.00 & 1.835  & 1.835 & 0.165 & 0.165 \\
1.20 & 1.704  & 1.704 & 0.296 & 0.296 \\
1.50 & 1.472  & 1.471 & 0.529 & 0.528 \\
1.70 & 1.335  & 1.335 & 0.665 & 0.666 \\
1.90 & 1.229  & 1.229 & 0.771 & 0.771 \\
20.00 & 1.000  & 1.000 & 1.000 & 1.000 \\
     & \multicolumn{4}{ c }{PNOF6}    \\
0.80 & 1.971 & 1.185 & 0.815 & 0.029 \\
1.00 & 1.942 & 1.197 & 0.803 & 0.058 \\
1.20 & 1.894  & 1.191 & 0.809 & 0.106 \\
1.50 & 1.771  & 1.150 & 0.850 & 0.230 \\
1.70 & 1.645	  & 1.110 & 0.891 & 0.355 \\ 
1.90 &  1.495 & 1.068 & 0.932 & 0.505 \\
20.00 & 1.000  & 1.000 & 1.000 & 1.000 \\ \hline
\end{tabular}

\end{table*}

The inclusion of interpair correlation also affects the occupation numbers of the corresponding NO (see table \ref{occ}). For small values of $R$ at the CASSCF level, the $a_g$ orbital remains almost doubly occupied along the PES. The $b_{2u}$ is doubly occupied for $\theta\ll90\dr$ and there is a smooth transition from these structures to the $\theta\gg90\dr$ ones in which the doubly occupied orbital is the $b_{3u}$. 
At $\theta=90\dr$ both orbitals become degenerate in terms of occupancies.  The PNOF5 bonding orbitals are almost doubly occupied along the PES while the antibonding ones remain almost unoccupied. No degeneracy is observed in this case. By including 
interpair electron correlation, PNOF6 NO and occupancies qualitatively mimic the CASSCF ones. It is worth noting that at $\theta=90\dr$ the $b_{3u}$
and $b_{2u}$ do not have exactly the same occupancy for most of the  values of $R$ shown in Fig. \ref{occ}. 
This might indicate that the interpair description is not fully recovered by PNOF6. The second solution shown in Fig. \ref{abs-170-190} as PNOF6(sol. 2), that becomes the minimum energy solution for certain values of $\theta$ when $R=1.70\AA$ and $1.90\AA$ and is the minimum solution found for larger values of $R$, presents perfect 
degeneracy in terms of occupation numbers of the $b_{3u}$
and $b_{2u}$ orbitals for all values of $\theta$ and $R$.\newline

The wrong coupling between spins located in diferent centers of the 
molecule is one of the causes for the failure of singlet-couplet geminal approaches 
to describe the \hh system. Jeszenszki \textit{et al.} have used the local spin analysis 
to show that the inclusion of triplet components in geminals improves the APSG results 
but spin contamination appears when the triplet component in the geminal becomes important.  
\begin{figure}
\includegraphics[scale=0.70]{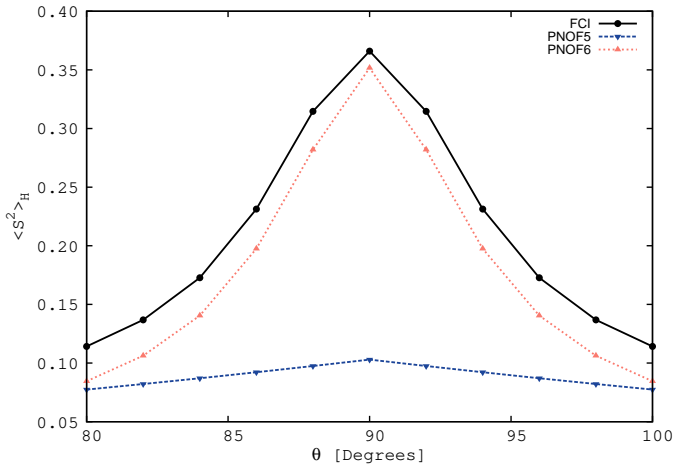}%

 \caption{\label{occu} Local spin values of one of the $H$ atom of \hh at $R=0.80\;\AA$ with respect to angle $\theta$.}
 \label{local}
\end{figure}
The local spin value of one $H$ atom of the \hh system is shown in Fig.~\ref{local}. 
As the system approaches the $D_{4h}$ symmetry, there is an increase of the diradical 
character of the system and the local spin on atom $H$ grows. PNOF5 cannot reproduce 
this trend and the local spin remains almost constant along the PES, while
PNOF6 local spin values in \hh are in good agreement with the FCI results.\newline

\begin{figure}
\includegraphics[scale=0.70]{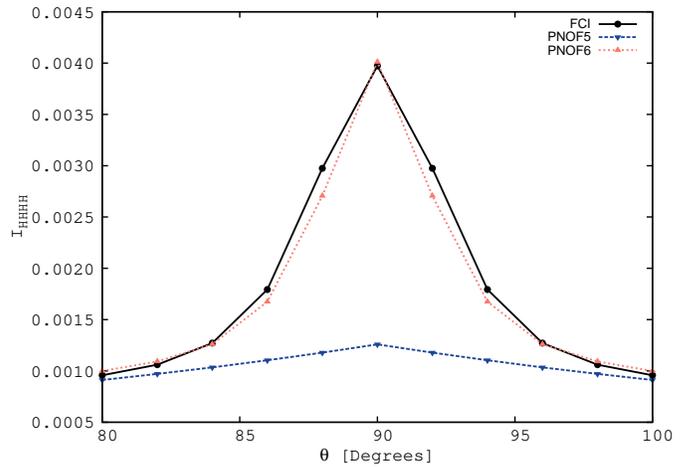}%

 \caption{Multicenter Giambiagi indices, Eq.~\ref{eq:multi}, along the $D_{2h}-D_{4h}$ transition
for $R=0.80\;\AA$ performed with PNOF5, PNOF6 and FCI natural orbitals.}
 \label{multi}
\end{figure}

Finally, let us examine the multicenter delocalization in the $D_{2h}$ to $D_{4h}$ transition
. Fig.~\ref{multi} shows that PNOF6 values closely follow the FCI ones and give a 
maximum electron delocalization in the $D_{4h}$ structures, whereas PNOF5 shows a rather
constant profile, clearly indicating its inability to delocalize the electron density along 
the \hh skeleton.\newline

\section{Conclusions}

The PES of the planar $D_{4h}/D_{2h}$ \hh model has been computed at several 
levels of theory. Single-reference methods show a spurious cusp at the $D_{4h}$
structure that thus far was attributed to nondynamic correlation.
PNOF5 (which affords a correct description
of molecular dissociation and other intrapair nondynamic correlation effects) also shows a
spurious cusp at $D_{4h}$, whereas PNOF6 provides a qualitatively correct 
description of this phenomenon. 

Since PNOF5 and PNOF6 mainly differ from each other by the inclusion of interpair
correlation, the factors responsible for the
spurious description of the $D_{4h}/D_{2h}$ \hh PES can
be narrowed down to missing \textit{interpair} nondynamic correlation effects.
Indeed, the inclusion of interpair correlation in the pairing-orbital NOFT ansatz 
is key to recover the delocalized orbitals picture, 
remove the spurious cusp in the PES and properly account for the coupling 
between the spins located at different centers. 
On the other hand, 
inclusion of more terms to fully account for electron correlation seems
to be needed to recover the smoothness of the curves at $R=1.70\AA$ and $1.90\AA$, to obtain quantitative results, and to recover the important correlation
effects that separate PNOF6 results from FCI.
We hope that this study will shed light on the effect of interpair electron correlation 
and pave the way to the development of new electronic structure methods within NOFT 
or methods based on geminal expansion of the wave function. 
Research in this direction is underway in our laboratory.\\

\begin{acknowledgments}
This research has been funded by the MINECO projects CTQ2012-38496-C05-01, CTQ2012-38496-C05-04 and CTQ2014-52525-P and the Basque Country Consolidated Group Project No. IT588-13.
We are grateful for the computational resources granted at the MareNostrum computer of the Barcelona Supercomputing Center
and technical and human support provided by SGI/IZO-SGIker UPV/EHU.
\end{acknowledgments}

\end{document}